\documentclass[oldversion,preprint]{aa}
\usepackage{graphicx}
\usepackage{txfonts}
\begin{document}

\title{Atmospheric turbulence in phase-referenced and wide-field interferometric images:}
\subtitle{Application to the SKA}

\titlerunning{Atmospheric turbulence in interferometric images}

\date{Submitted to A\&A on 2009 December 4; accepted on 2010 April 2.}

\author{I. Mart\'i-Vidal\inst{1,2}
        \and J.C. Guirado\inst{2}
        \and S. Jim\'enez-Monferrer\inst{2} 
        \and J.M. Marcaide\inst{2}}
\institute{Max-Planck-Institut f\"ur Radiastronomie, Auf dem H\"ugel 69, D-53121 Bonn (Germany) 
           \email{imartiv@mpifr.de}
           \and 
           Dpt. Astronomia i Astrof\'isica, Universitat de Val\`encia, Dr. Moliner 50, 
           E-46100 Burjassot (Spain)}

\abstract{
Phase referencing is a standard calibration procedure in radio interferometry. It allows to 
detect weak sources by using quasi-simultaneous observations of closeby sources acting as 
calibrators. Therefore, it is assumed that, for each antenna, the optical paths of the 
signals from both sources are similar. However, atmospheric turbulence 
may introduce strong differences in the optical paths of the signals and affect, or even 
waste, phase referencing for cases of relatively large calibrator-to-target separations and/or 
bad weather. The situation is similar in wide-field observations, since the random
deformations of the images, mostly caused by atmospheric turbulence, 
have essentially the same origin as the random astrometric variations of phase-referenced 
sources with respect to the phase center of their calibrators. In this paper, we present the 
results of a Monte Carlo study of the astrometric precision and sensitivity of an 
interferometric array (a realization of the Square Kilometre Array, SKA) in phase-referenced 
and wide-field observations. These simulations can be extrapolated to other arrays by applying the
corresponding corrections. We consider several effects from the turbulent atmosphere (i.e., 
ionosphere and wet component of the troposphere) and also from the antenna receivers. We 
study the changes in dynamic range and astrometric precision as a function of observing frequency, 
source separation, and strength of the turbulence. We find that, for frequencies 
between 1 and 10\,GHz, it is possible to obtain images with high fidelity, although the atmosphere 
strongly limits the sensitivity of the instrument compared to the case with no atmosphere. Outside 
this frequency window, the dynamic range of the images and the accuracy of the source positions 
decrease. 
We also find that, even if a good model of the atmospheric turbulence (with an accuracy of 
99\%) is used in the imaging, residual effects from the turbulence can still limit the dynamic 
ranges of deep, high-contrast ($10^5 - 10^6$), images.
}


 \maketitle

\section{Introduction}

It is well-known that ground-based astronomical observations are affected by the atmosphere. 
Changes in the atmospheric opacity produce a bias in the source flux density, while changes in the
refraction index distort the shape of the electromagnetic frontwave of the source. Such a distortion 
translates into a deformation of the observed source structure and/or a variation of the relative 
positions of all sources observed in a given field. In the case of astronomical devices based on 
interferometry, atmospheric effects can be well modelled if the atmosphere above each 
element of the interferometer (hereafter, {\em station}) remains unchanged over the whole portion 
of the sky being observed. In such cases, the observed visibilities can be calibrated using 
{\em station-based} algorithms, which are relatively simple and computationally inexpensive 
(e.g. Readhead \& Wilkinson \cite{Readhead1978}).

However, when the spatial variations of the atmosphere are significant within the observed 
portion of the sky, as it happens if there is atmospheric turbulence, the opacity and 
dispersive effects cannot be modeled as a single time-dependent station-based complex gain 
over the field of view.  Unless more complicated calibration algorithms are used 
(e.g., van der Tol, Jeffs, \& van der Veen \cite{vdT07}), the effect of these errors on the 
image are difficult to correct. In this paper, we report on a study of the effects 
that a turbulent atmosphere may introduce in interferometric observations. We focus our 
study on the effects produced by turbulence in the dynamic range and astrometric accuracy 
after a phase-referenced calibration between a strong (calibrator) source and a weak source, 
located a few degrees away. This study is numerically equivalent to the study of the 
deformation of a wide-field interferometric image at any point located at a given distance 
from the center of the field (i.e., the {\em phase center} of the image). In both cases, 
the phases introduced by the atmosphere in the signal of each antenna for the different 
pointing directions are the same, so the effects of the atmosphere in Fourier space (and 
therefore on the sky plane) will also be the same.

The results here reported are an extended version of those previously reported in 
the SKA memo by Mart\'i-Vidal et al. (\cite{MartiVidal2009}). In the next section, we describe 
the details of the array distribution used, as well as the characteristics of the simulated 
observations. In Sect. \ref{NoiseModel}, we describe how the noise from the atmosphere and the 
receivers was added to the visibilities and in Sect. \ref{algorithm} describe the procedures 
followed in our Monte Carlo analysis. In Sect. \ref{Results}, we present the main results 
obtained; in Sect. \ref{Summary}, we summarize our conclusions.

\section{Array geometry and sensitivity}

\begin{figure*}[ht!]
\centering
\includegraphics[width=18cm,angle=0]{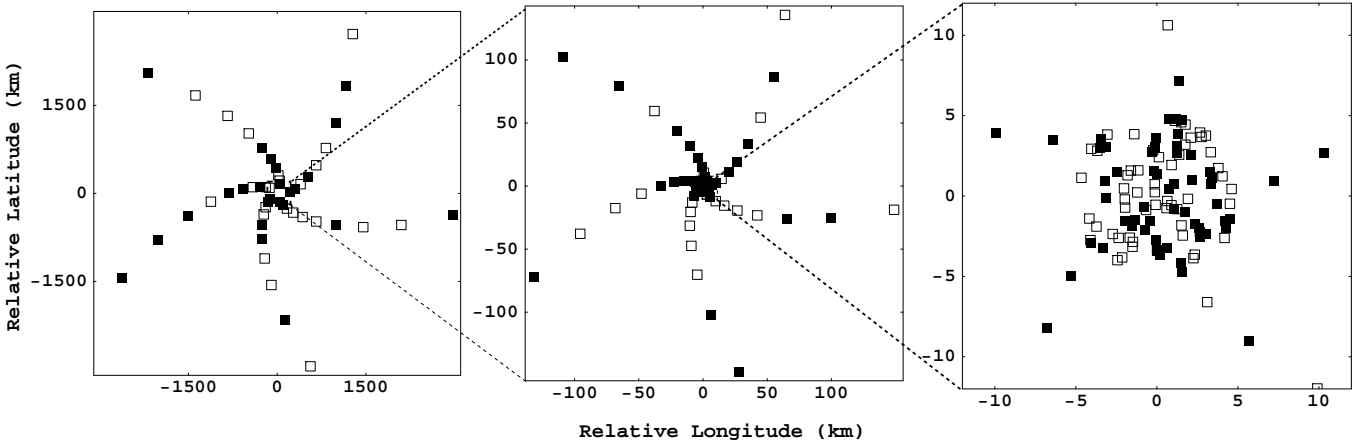}
\caption{Array distribution used in our simulations (empty and filled squares). Axes are relative 
Longitude (horizontal axis) and Latitude (vertical axis) in km. Left, the whole array. Center, a 
zoom to the core. Right, a zoom to the inner core. The stations marked with empty squares were 
removed from the array in a second run of our simulations, to check the dependence of the results 
on different array distributions (see text).}
\label{ARRAY}
\end{figure*}

We simulated an interferometric array similar to the planned station distribution of the 
Square Kilometre Array (SKA). We simulated a total of 200 stations distributed in the following 
way: 50\% are randomly distributed within a circle of 5\,km radius (inner core); 25\% are 
distributed outside this circle up to a distance of 150\,km (core), following 5 equiangular spiral 
arms; the remaining antennae are distributed following the same spiral arms, but up to a distance 
of 3000\,km from the inner core. This array distribution is similar to that used in Vir Lal, 
Lobanov \& Jim\'enez-Monferrer (\cite{VLJ09}). The curvature of the Earth surface was taken 
into account in our simulations. 
We show the resulting array distribution in Fig. \ref{ARRAY}. We also repeated all the 
simulations here reported, but subtracting a subset of 100 (randomly selected) stations from the array, 
to check the sensitivity of the main conclusions of this paper on different array distributions (see
Appendix \ref{OtherArrays}). We also show in Fig. \ref{ARRAY} the modified array after subtracting 
the 100 stations.

\subsection{Sensitivity and bandwidth}

We simulated interferometric observations using 16 different frequencies, which span in 
logarithmic bins from 150\,MHz to 24\,GHz (this is the theoretical frequency window of the SKA). 
According to Jones (\cite{J04}), the maximum observing bandwidth of the SKA will be around 25\% 
of the central observing frequency (up to a maximum bandwidth of 4\,GHz for all frequencies 
above 16\,GHz). This (maximum) frequency-dependent bandwidth translates in our simulations 
into a changing sensitivity of the SKA as a function of frequency.

\begin{figure}[ht!]
\centering
\includegraphics[width=9cm,angle=0]{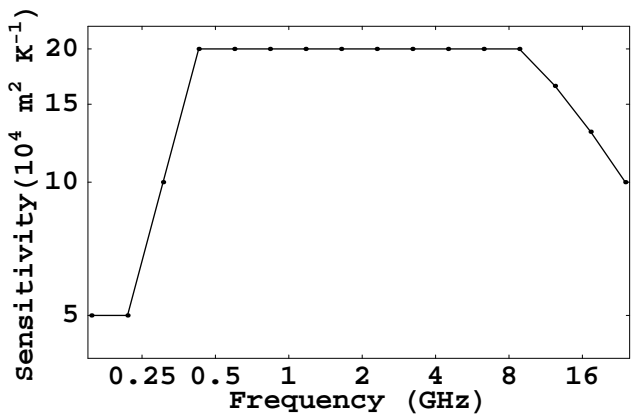}
\caption{\small Station sensitivities (i.e., effective areas over system temperatures) used in our 
simulations.}
\label{SENSITIVITY}
\end{figure}

The sensitivities of the simulated stations were also chosen to be similar to those of the SKA, 
which were taken from Jones (\cite{J04}). These values are set for an elevation of 45 degrees 
and differ from those given in Schilizzi et al. (\cite{SAC07}), but the use of the values given in 
Schilizzi et al. (\cite{SAC07}), instead, does not affect the main conclusions of this paper. We 
interpolated the sensitivities given in Table 1 of Jones (\cite{J04}) to the frequencies used 
in our simulations. In Fig. \ref{SENSITIVITY} we show the station sensitivities used.   

\subsection{Source position}

We set the target source coordinates at the zenith of the array center and the calibrator at 
an hour angle of 0 degrees, also with respect to the array center. This position of the sources 
minimizes the optical paths of 
the signals through the atmosphere (since both, source and calibrator, are at maximum elevations), 
thus enhancing the quality of the phase-referenced observations. The results given in this paper 
should be interpreted according to this issue. 

If the source would be located far from the zenith,
the mapping function of the tropospheric delay and the finite width of the ionosphere would 
increase the effect of turbulence on the phase-referenced visibilities of the target.
Additionally, the uv coverage of the interferometer would have shorter projected baselines in 
declination, thus decreasing the synthesized resolution in declination.
Therefore, we want to stress that setting the sources at maximum elevations is a key limiting 
factor of the simulations here reported, especially if they are to be compared to real 
observations.

\section{Noise model}
\label{NoiseModel}

We simulated phase-referencing observations in the following way: we assumed that the calibrator 
source is sufficiently strong to allow for a perfect antenna-gain calibration at its location; we 
then determined the image of the target source by computing the differential antenna-gain errors 
expected at the target location. Therefore, under the effect of atmospheric turbulence, these results 
depend on the calibrator-to-target separation.

We implemented two kinds of atmospheric turbulence. The first turbulence was associated 
to the ionosphere (the free electron content, which introduces dispersion in the radiation) and the 
other turbulence was associated to the wet troposphere (the water vapour, close to the earth 
surface, which is in a state of no thermodynamic equilibrium). The effect of ionospheric turbulence 
on the signal phase varies as $\nu^{-1}$, affecting the low-frequency observations; the effect of the 
wet troposphere on the phase varies as $\nu$, affecting the high-frequency observations. 
The dry troposphere (which is more homogeneously distributed over each station 
than the wet troposphere) was not considered in our simulations, since it can be easily 
modeled and removed from the data to a level lower than the effects coming from the water vapor and 
the ionosphere. Models of the turbulence from the ionosphere and troposphere can be found in many 
publications (e.g., Thomson, Moran, \& Swenson \cite{TMS91}). Here, it is suffice to say that this 
turbulence follows a Kolmogorov distribution. This distribution has a phase structure function given by

\begin{equation}
D_{\phi}(\theta) = ~ <\left(\Phi(\theta_0) - \Phi(\theta_0 + \theta)\right)^2> ~ \propto ~ \theta^{5/3}
\end{equation}

\noindent where $\Phi(\theta_0)$ is the phase added by the turbulent screen to the signal of a source 
located at $\theta_0$. The brackets $<...>$ represent averaging over all pointing directions located at 
a distance $\theta$ from the point at $\theta_0$. The Kolmogorov distribution is fractal-like, so both, 
ionosphere and wet troposphere, have essentially the same phase distribution, despite of a global scaling 
factor between them. 

The global factors for both distributions (ionosphere and troposphere) were computed according to the 
typical values of ionospheric and tropospheric conditions. For the ionosphere, the Fried length (i.e., 
distance in the ionosphere for which the structure function rises to 1\,rad$^2$) was set to 3\,km at 
100\,MHz. For the wet troposphere, we set the parameter $C^2_n L$ (i.e., the integral of the profile of 
$C^2_n$ along the zenith direction) to $10^{-11}$\,m$^{1/3}$ (vid. Eq. 13.100 and Table 13.2 of Thomson, 
Moran, \& Swenson \cite{TMS91}); this value translates into a Fried length of 3\,km for a frequency of 
$\sim22$\,GHz. Since the Kolmogorov distribution is self-similar, it is possible to adapt the results 
here reported to any other atmospheric conditions (see Sect. \ref{ChangingR0}), just accordingly scaling 
the source separation to the Fried length of the ionosphere (for low-frequency observations) or the wet 
troposphere (for high-frequency observations). We must notice that the self-similarity of the tropospheric 
turbulence does not hold for very large scales (the typical baseline lengths in VLBI observations), 
since there is a saturation in the power spectrum of the distribution (see, e.g., Thomson, Moran, 
\& Swenson \cite{TMS91}). However, this is not important in our analysis, since we did not use the absolute 
phase of the signal coming from a given direction in the sky, but computed the differential effects 
at each station from two different (closeby) directions, which depend on short-scale turbulence. 
Therefore, the saturation of tropospheric turbulence at large scales does not affect our results.

\begin{figure}[ht!]
\centering
\includegraphics[width=9cm,angle=0]{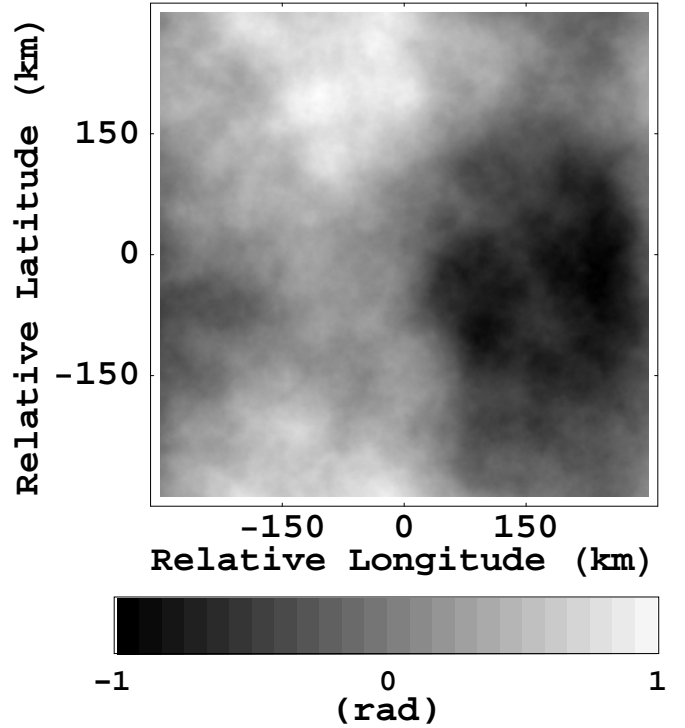}
\caption{Example of a turbulent phase screen with Kolmogorov statistics. The grey scale shows variations 
of optical-path phases, normalized between $-1$ and 1 radians. The final values of the phases depend 
(given the self-similarity of the distribution) on a global factor related to the observing frequency 
and the ionospheric and/or wet tropospheric conditions.}
\label{TURBULENCE}
\end{figure}

We computed the differential effects from the turbulent atmosphere in two ways. For the antennas of the 
core (within the central 300\,km) we generated synthetic phase screens for the ionosphere and troposphere. 
We show an example of one such screen in Fig. \ref{TURBULENCE}. We notice that this figure could represent 
either ionospheric or tropospheric turbulence in our modeling, just by scaling the screen 
by the corresponding factor. Two different screens were generated in each Monte Carlo simulation. 
The screen for simulating the ionosphere was put 
at a height of 300\,km and the screen for simulating the troposphere was put at a height of 5\,km. For the 
antennas out of the core, we computed the term $\Phi(\theta_0) - \Phi(\theta_0 + \theta)$ separately. We 
proceeded this way (i.e., we generated a phase screen only for the core antennas, thus without generating 
a much larger screen for the whole array), because the distances between stations out of the core are large 
enough to ensure that the cross-correlation of turbulence above different stations is negligible 
compared to the correlation between those on the calibrator and target source for the same station. 
This numerical strategy also speeded up our simulations.

It must be noticed that we did not introduce any time evolution of the turbulent phase screens in our 
simulations. Any evolution of the turbulence could dramatically affect the observations if the 
acquisition times were larger than the {\em coherence time} of the signal, which depends on the evolution of 
the turbulence and the observing frequency. However, for snapshot-like observations, of the order of a 
fraction of a minute or so, we could consider, as a good first approximation, a constant turbulence 
phase screen.

Noise from the receivers was added to our model by generating a random Gaussian noise in the real and 
imaginary parts of the visibilities. The mean deviation, $\sigma$, of the Gaussian noise added to the 
visibilities was (e.g., Thomson, Moran, \& Swenson \cite{TMS91}, Eq. 6.43):

\begin{equation}
\sigma = \frac{\sqrt{2}\,k}{\eta_Q\sqrt{\Delta\nu\,\Delta t}}\,\frac{1}{S_A}
\end{equation}

\noindent where $k$ is the Boltzmann constant, $\eta_Q$ is the relative loss of signal due to the correlator 
quantization (we used $\eta_Q = 0.5$), $\Delta\nu$ is the observing bandwidth, $\Delta t$ is the observing 
time, and $S_A$ is the sensitivity of the stations (collecting area over system temperature, shown in 
Fig. \ref{SENSITIVITY}).

\section{Estimate of dynamic range and astrometric precision}
\label{algorithm}

We simulated different sets of phase-referenced observations. In all cases, the observations were snapshots 
with a duration $t_0 = 60$\,s. Longer observing times, $t$, would, in principle, increase the dynamic ranges 
and astrometric precisions shown in all the following sections as $\sqrt{t/t_0}$, as long as the changing 
atmosphere (and, therefore, the changing source positions and shapes) would not introduce important smearing 
effects in the images after the combination of all visibilities. 

In a first run of simulations, we generated visibilities of targets with flux densities of 0.1, 1, and 
10\,$\mu$Jy with a separation of 5 degrees between target and calibrator. A total of 1500 simulations were 
performed for each flux density and frequency. We used such a large separation between calibrator and target, 
because these simulations of phase-referenced observations can also be applied 
to the study of deformations of wide-field images under the effects of a turbulent atmosphere. 

In a second run of simulations, we studied the effects of the atmosphere as a function of calibrator-to-target 
separation. For that purpouse, we simulated 1500 observations at 1420\,MHz (i.e., the Hydrogen line) of a 
source with 1\,$\mu$Jy for different separations from the calibrator (2, 3, 4, 5, and 6 degrees). 

In a third run of simulations, we used only one Kolmogorov screen (which can represent either ionospheric or 
tropospheric turbulence, depending on the observing frequency) with different Fried lengths, to study 
the scalability of the simulations for different source separations and/or atmospheric conditions. 

\begin{figure}[ht!]
\centering
\includegraphics[width=9cm,angle=0]{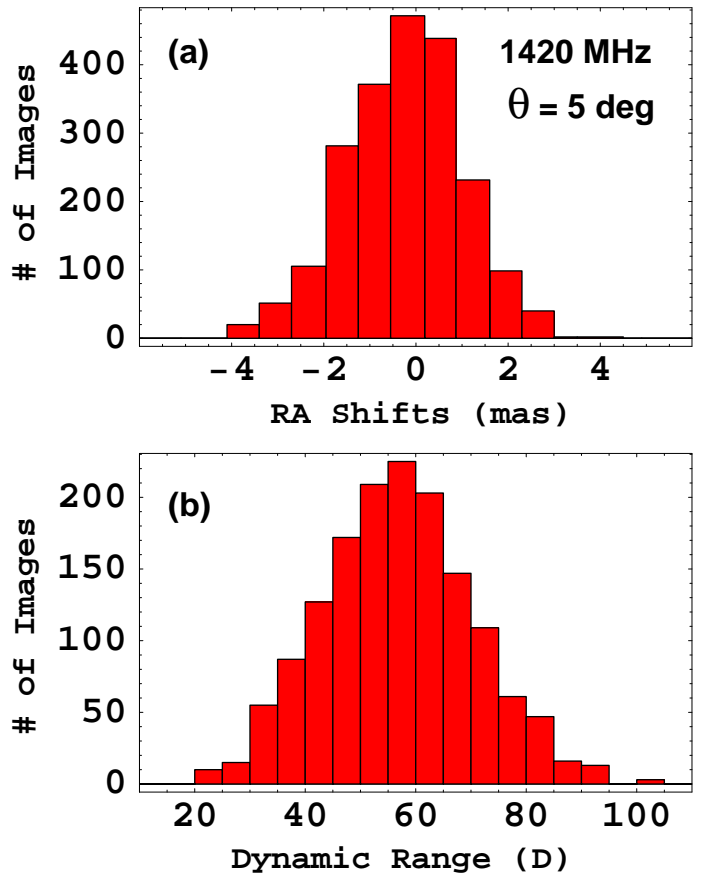}
\caption{Distribution of right ascension shifts (a) and dynamic ranges (b) of 1500 snapshot 
phase-referenced images, simulated at 1420\,MHz, for a 1\,$\mu$Jy target source located at 5\,deg from 
its calibrator. The calibrator source is located at an hour angle of 0 and the target source is 
located at the zenith of the array center.}
\label{HISTOGRAMS}
\end{figure}

In all these simulations, we added the noise from the atmosphere and the noise from the 
receivers. 
For each simulated phase-reference image, obtained by applying uniform weighting to the visibilities, the 
brightness peak was found and the corresponding point source was subtrated from the visibilities. For the 
substraction of the point source, the brightness peak was shifted to the phase center of the image by 
multiplying the visibilities by the corresponding plane-wave factor in Fourier space. Then, the flux 
density of the point source was estimated as the average of the real part of the resulting visibilities, and
the resulting point-source model was substracted from the data. Afterwards, a Fourier inversion of the new 
visibilities resulted in the image of residuals, from which the root-mean-square (rms) of all the 
pixels was computed. 
On the one hand, the deviation of the brightness peak with respect to the image center was taken as 
the astrometry error of that image. On the other hand, the source peak divided by the rms of the residuals 
was taken as the dynamic range. In Fig. \ref{HISTOGRAMS} we show the distribution of astrometric deviations 
and dynamic ranges for the case of a target source of 1\,$\mu$Jy observed at 1420\,MHz (which corresponds 
to an interferometric beam of $\sim$13\,mas) located at 5 degrees from the calibrator. Once the distributions 
like those shown in Fig. \ref{HISTOGRAMS} were obtained, we computed the standard deviation of 
astrometric corrections and the mean value of dynamic ranges for each source flux density, frequency, and 
separation. The first quantity was our estimate of the astrometric uncertainty, and the second quantity was 
an estimate of the achievable dynamic range.

\section{Results}
\label{Results}

\subsection{Observing frequency and signal decoherence}
\label{WITHAT}

If the atmospheric turbulence is not taken into account and only the noise from the receivers 
is added to the visibilities, our simulations reproduce the dynamic ranges 
given by Eq. 6.53 of Thomson, Moran, \& Swenson (\cite{TMS91}), as expected.
Additionally, the noise from the receivers does not introduce considerable changes in the source position 
of the phase-referenced images (changes of the order of 10\,$\mu$as or lower). 

\begin{figure*}[ht!]
\centering
\includegraphics[width=17cm]{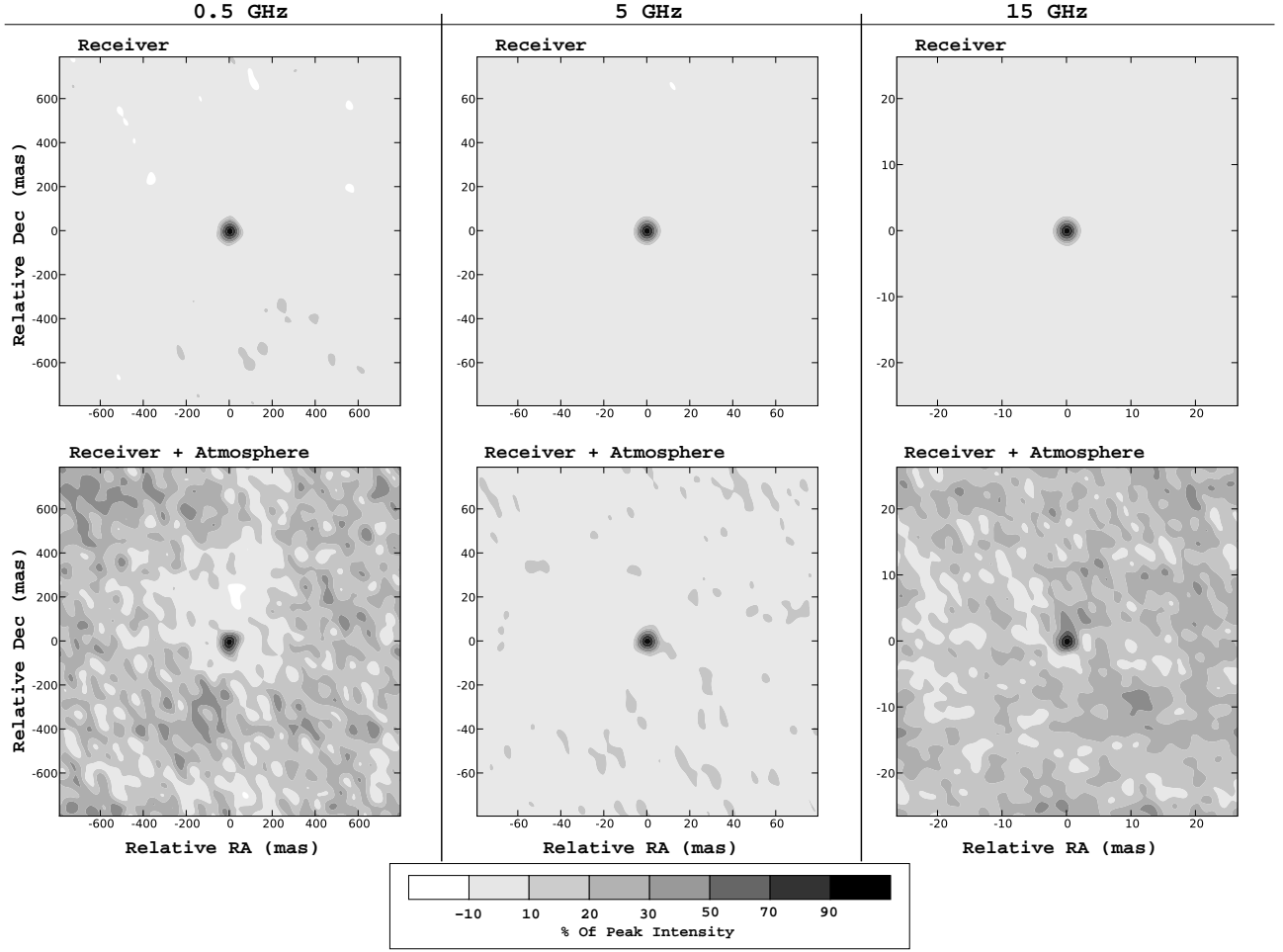}
\caption{Simulated phase-reference images of a 1\,$\mu$Jy source located at 5\,deg from its calibrator. 
Noise from the receiving system (top) and, additionally, noise due to the turbulent atmosphere (bottom) 
have been added to the visibilities.}
\label{Mapas}
\end{figure*}

When the turbulent ionosphere 
and wet troposphere are added to the simulations, the dynamic range of the images is notably affected, 
especially at low ($<1\,GHz$) and high ($>10\,GHz$) frequencies. In Fig. \ref{Mapas}, we show 
phase-referenced images of a 1\,$\mu$Jy source, 
located at 5\,deg. from its calibrator, observed at 0.5, 
5, and 15\,GHz. It can be 
readily seen that the addition of effects coming from the atmospheric turbulence maps into an important 
extra noise in the images at 0.5 and 15\,GHz, but not at 5\,GHz.

\begin{figure}[ht!]
\centering
\includegraphics[width=9cm,angle=0]{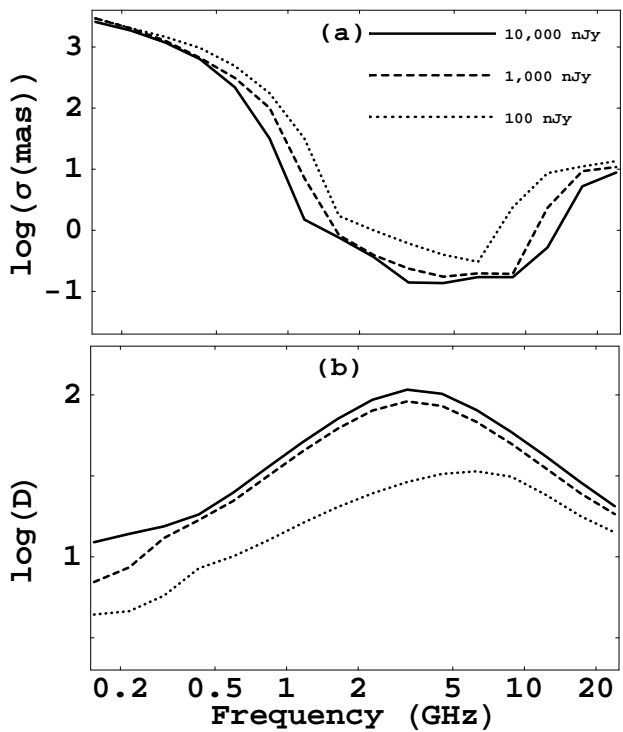}
\caption{Astrometric accuracy (a) and dynamic range (b) as a function of frequency, for the case with 
atmospheric turbulence and a separation of 5 degrees between calibrator and target. The calibrator 
source is located at an hour angle of 0 and the target source is located at the zenith of the array 
center. Different lines 
correspond to different target source flux densities (10\,$\mu$Jy, continuous line; 1\,$\mu$Jy, dashed 
line; 0.1\,$\mu$Jy, dotted line).}
\label{WITHATMOSPHERE}
\end{figure}

Following the algorithm described in Sect. \ref{algorithm}, we obtained the astrometric 
uncertainties and dynamic ranges shown in Fig. \ref{WITHATMOSPHERE}. For very low frequencies (below 
$\sim$500\,MHz) the ionosphere prevents a clear and precise detection of all sources, no matter their flux 
densities. For higher frequencies, the astrometric uncertainty decreases notably (mainly because of the 
dependence of ionospheric effects with $\nu^{-1}$) and gets limited only by diffraction and sensitivity 
between 1 and 10\,GHz (this frequency window slightly depends on the source flux density, as it can be 
seen in the figure). For higher frequencies, the wet troposphere begins to affect the astrometric 
uncertainty, which rises up to around 10\,mas for the highest frequencies. We find that the best 
astrometric accuracy, at least for reasonably well-detected sources, is achieved for frequencies around 
4\,GHz. This is where the ionospheric and (wet) tropospheric components are roughly equal.

The dynamic range of the phase-referenced images is highly limited by the atmosphere. When the atmosphere 
adds noise to the visibility phases, there is an extra rms added to the residual images, 
which depends on the visibility amplitudes, thus limiting the achievable dynamic range no matter the 
flux density of the source; that is, if the source flux density is higher, the noise of the image 
will also be higher. This limitation is, of course, more important for the brightest sources. In our 
case, the brightest source has a flux density of 10\,$\mu$Jy. For this source, the maximum dynamic range 
achieved is only 110, which is $\sim30$ times smaller than the dynamic range that would be obtained without the 
atmosphere. This situation can be also understood in another way: the rms of the final image is divided into two 
components, which are added in quadrature. One component, $\sigma_{th}$, comes from the receiver noise 
and is independent of the source flux density. The other component, $\sigma_{at}$, comes from the 
atmospheric refraction and is equal to a percentage of the source flux density ($\sigma_{at} = K_r\,S$, 
where $S$ is the source flux density and $K_r$ depends on the atmospheric refraction). Hence, 
the dynamic range, $D$, is

\begin{equation}
D = \frac{S}{\sqrt{\sigma_{th}^2+\sigma_{at}^2}} = \frac{S}{\sqrt{\sigma_{th}^2+K_r^2\,S^2}}
\label{SNRth}
\end{equation}

For large flux densities ($S>>\sigma_{th}$), the achievable dynamic range will saturate to a value 
dependent on the atmospheric conditions (i.e., $D \rightarrow 1/K_r$) and independent on the source flux 
density and the sensitivity of the stations. 
As seen in Fig. \ref{WITHATMOSPHERE}, despite of the large difference between the simulated flux 
densities (1 and 10 $\mu$Jy), the dynamic range saturates to a value of $\sim$100.

\subsection{Angular separation and signal decoherence}

\begin{figure}[ht!]
\centering
\includegraphics[width=9cm,angle=0]{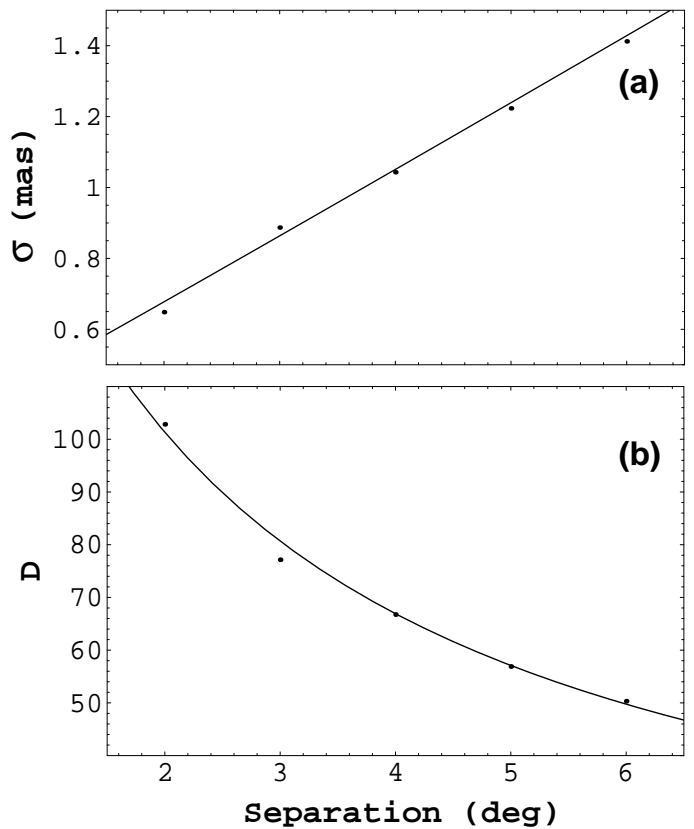}
\caption{Dots: simulated astrometric accuracy (a) and dynamic range (b) as a function of angular
separation between calibrator and source, for observations at 1420\,MHz and a target flux density
of 1\,$\mu$Jy. Lines: proposed phenomenological models.}
\label{WITHDISTANCE}
\end{figure}

The results shown in the previous subsection correspond to a separation of 5 degrees between source and 
calibrator. These results change when the angular separation changes. We computed astrometric 
uncertainties and dynamic ranges for a source with a flux densitity of 1\,$\mu$Jy located at 2, 3, 4, 
5, and 6 degrees from its calibrator. Noise from the atmosphere and the receivers was taken 
into account in these simulations. We used an observing frequency of 1420\,MHz (the Hydrogen line) 
which is inside the frequency window where the atmospheric effects are minimised. Therefore, all the 
astrometric errors derived were small (of the order of a few mas), allowing us to use image sizes small 
enough to sample the beam with more pixels ($\sim$30 pixels) using a grid of 1024$\times$1024 pixels. 
This fine gridding of the beam allowed for a more accurate determination of the location of the image 
peak and, therefore, a better estimate of the astrometric error. The results obtained are shown in Fig. 
\ref{WITHDISTANCE}. In that figure, we also plot two analytical (phenomenological) models for the 
estimate of the increase of astrometric uncertainty and the loss of dynamic range (i.e., degree of 
signal decoherence) as a function of angular separation. On the one hand, the phenomenological model 
proposed for the estimate of loss of dynamic range is:

\begin{equation}
D = \frac{D_0}{\sqrt{1 + k^2\,D_0^2\,\theta^{\beta}}},
\label{DynRanEq}
\end{equation}

\vspace{0.5cm}
\noindent where $D$ is the dynamic range, $\theta$ is the angular separation between target and calibrator, 
$D_0$ is the dynamic range without atmosphere (i.e., when the calibrator-to-target separation, $\theta$, 
tends to 0), and $k$ and $\beta$ are two parameters related to the atmospheric conditions, 
source flux density, and observing frequency. This equation is just Eq. \ref{SNRth}, but setting 
$K_r = k\,\theta^{\beta/2}$. 

As it can be seen, this model fits well to the simulations. We obtain
$k = (2.5\pm0.2)\times10^{-5}$\,deg$^{-\beta}$ and $\beta = 1.51\pm0.06$.

On the other hand, the proposed phenomenological model for the increase of astrometric uncertainty is:

\begin{equation}
\sigma = \frac{\sigma_0}{D_0}\,(1 + k'\,\theta^{\beta'}),
\label{SigEq}
\end{equation}

\vspace{0.5cm}
\noindent where $\sigma$ is the astrometric uncertainty, $\sigma_0$ is the diffraction limit (i.e., half 
the size of the beam), and $k'$ and $\beta'$ are two parameters also related to the 
atmospheric conditions, source flux density, and observing frequency. We fit $k' = 0.57\pm0.04$\,deg$^{-\beta'}$ 
and $\beta' = 1.02\pm0.05$. From a different approach in the treatment of the noise from the atmosphere
and limited to VLBA and EVN arrays, Pradel, Charlot, \& Lestrade (\cite{Pradel}) obtained a similar dependence of $\sigma$ with $\theta$.

We notice that if we change $D_0$ by $D$ in Eq. \ref{SigEq}, the new fitted $k'$ and $\beta'$ are
1.03$\pm$0.09 and 0.07$\pm$0.06, respectively. This new value of $\beta'$ is compatible with 
zero. In other words, the diffraction limit divided by the dynamic range of the image is 
an excellent estimator of the astrometric uncertainty, at least for the range of simulated 
calibrator-to-target separations at 1.4\,GHz (which falls within the frequency window where the 
atmospheric effects are minimized).

For calibrator-to-target separations larger than $\sim$6 degrees, the situation changes. 
We simulated phase-referenced images for calibrator-to-target separations up to 12 
degrees, and found that the model of dynamic range given by Eq. \ref{DynRanEq} is still valid, 
but the astrometric uncertainty increases faster, with $\beta' = 2.38\pm0.08$ 
($\beta' = 1.38$, if we change $D_0$ by $D$ in Eq. \ref{DynRanEq}). This 
last $\beta'$ fits well to the astrometric uncertainties for large source separations, 
but the fit is worse for separations smaller than 5-6 degrees.

\subsection{Scalability of the results and use of turbulence models in the data calibration}
\label{ChangingR0}

\begin{figure}[ht!]
\centering
\includegraphics[width=9cm,angle=0]{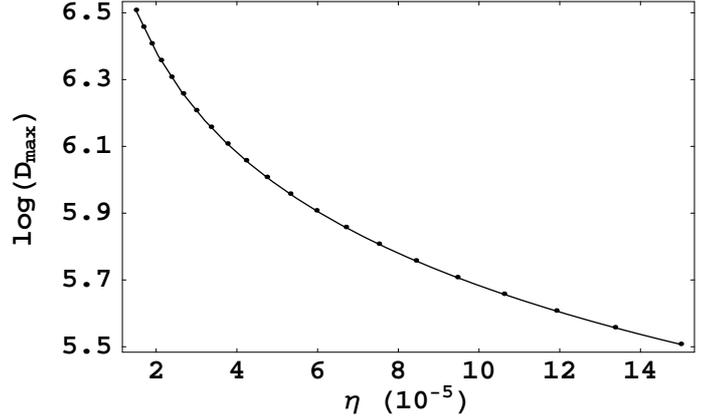}
\caption{Maximum dynamic range as a function of $\eta$ (see Eq. \ref{NormDist}) and for a 1-scan snapshot 
of 60\,s.}
\label{WithR0}
\end{figure}

In the previous subsections, we report on the effects of atmospheric turbulence in phase-referenced 
(and wide-field) interferometric images using fixed values for the Fried lengths of the Kolmogorov 
distributions of the ionosphere and wet troposphere. Since the Kolmogorov distribution is self-similar, 
the results reported can be scaled and adapted to other atmospheric conditions. Indeed, these 
simulations can also be used to estimate the limiting dynamic range and astrometric uncertainty if an 
a priori model of the tropospheric and/or ionospheric turbulence is used in the imaging. In these 
cases, the effective Fried length, $r_{ef}$, to compare to our simulations can be estimated as 

\begin{equation}
r_{ef} = r_0 \langle \frac{\phi_{mod}}{|\phi - \phi_{mod}|} \rangle
\end{equation}

\noindent where $r_0$ is the Fried length of the real turbulence and the other factor is related to 
the fractional precision of the turbulence model: $\phi_{mod}$ is the phase computed from the turbulence 
model at a given point in the sky and $\phi$ is that corresponding to the real turbulence; 
the brackets $<...>$ represent averaging over the field of view. 
If the a priori model of the ionospheric electron distribution is accurate to a given precision level, 
the effective Fried length to use for the ionosphere will be that corresponding 
to the {\em residual} turbulence (i.e., the difference between the model and the real turbulence). For 
instance, in the case of a model of the ionospheric electron content with a 99\% accuracy, the 
effective Fried length will be $r_{ef} = 100\,r_0$; if the accuracy increases to 99.9\%, 
$r_{ef} = 1000\,r_0$. In Fig. \ref{WithR0}, we show $D_{max}$, the maximum dynamic range (i.e., for a 
source with an infinite flux density, so $\sigma_{th} =0$ in Eq. \ref{SNRth}) as a function of 
parameter $\eta$, which we define as

\begin{equation}
\eta = \frac{h\,\sin{\theta}}{r_{ef}}
\label{NormDist}
\end{equation}

In this equation, $h$ is the height of the phase screen and $\theta$ is the calibrator-to-target separation
(or half the size of the wide-field image). Equation \ref{NormDist} can be used, together with 
Fig. \ref{WithR0}, to compute the maximum achievable dynamic range for many different combinations of source 
separations, atmospheric conditions, and observing frequencies ($r_{ef} \propto \nu$ for the ionosphere and 
$r_{ef} \propto \nu^{-1}$ for the trosposphere). We notice, however, that Fig. \ref{WithR0} has been generated 
using only one Kolmogorov screen, so it is applicable to ionospheric dispersion (for low frequencies) or 
tropospheric dispersion (for high frequencies), but not to a situation where ionospheric and tropospheric 
effects are similar. In these cases, and as a first approximation, we could set 

\begin{equation}
\eta = \frac{h_{ion}\,\sin{\theta}}{\sqrt{r_{ion}^2+\left(\frac{h_{ion}}{h_{trop}}r_{trop}\right)^2}}
\end{equation}

\noindent where $r_{ion}$ and $r_{trop}$ are the (effective) Fried lengths of the ionosphere and troposphere, 
respectively, and $h_{ion}$ and $h_{trop}$ are the heights of each phase screen. 
Also shown in Fig. \ref{WithR0} is the fitting model 

\begin{equation}
D_{max} = D_1\,\eta^{\beta_1}
\label{DmNorm}
\end{equation}

\noindent where $D_1 = 48.38\pm 0.10$ and $\beta_1 = -1.000\pm 0.002$.
Figure \ref{WithR0} indicates how difficult is to obtain a high-contrast image with a wide-angle coverage 
at very low (or high) frequencies. For instance, an image of $10\times10$ degrees with a dynamic range 
of $\sim$$10^6$ at a frequency of 500\,MHz (which translates into $\eta \sim 5\times10^{-5}$) would require, 
for the ionospheric screen used in the previous subsections, a model of the ionospheric turbulence 
distribution with an uncertainty lower than $3\times10^{-5}$ (observing time of 1 minute).
If the observing time increases to 10 hours, the minimum required uncertainty in the model of the 
ionosphere would increase to $7\times10^{-4}$, but such a high precision should be kept during the 
whole set of observations. If the dynamic range decreases to $10^5$, the required minimum uncertainty
for the ionosphere model would decrease to $7\times10^{-3}$ (i.e., 0.7\%) for an observing time of 10 
hours.

\begin{figure}[ht!]
\centering
\includegraphics[width=9cm,angle=0]{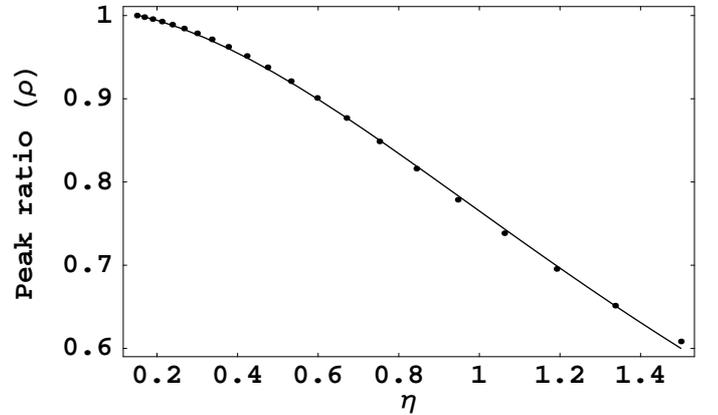}
\caption{Peak flux density of the phase-referenced image, relative to the real peak flux density of the 
source, as a function of $\eta$ (see Eq. \ref{NormDist}) and for a 1-scan snapshot of 60\,s.}
\label{PeakRatio}
\end{figure}

\begin{figure}[ht!]
\centering
\includegraphics[width=9cm,angle=0]{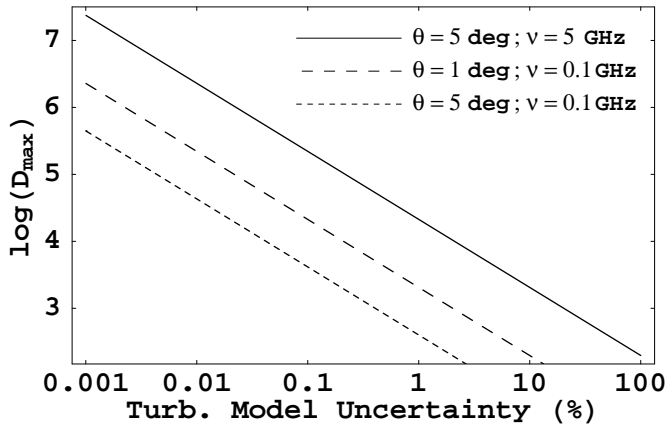}
\caption{Maximum dynamic range as a function of the uncertainty in the model of the atmospheric turbulence
for a 1-scan snapshot of 60\,s.}
\label{DmaxUncert}
\end{figure}

For a more comprehensive representation of our results, 
we show in Fig. \ref{DmaxUncert} how the achievable dynamic range (computed from Eq. \ref{DmNorm}) depends 
on the uncertainty in the model of atmospheric turbulence used in the data calibration. We show this 
relationship for 
different observing frequencies and calibrator-to-target separations. For instance, a dynamic range $10^4$
in observations at 100\,MHz for a calibrator-to-target separation of 1 deg. (i.e., the same for a wide-field
image of $2\times2$ deg.) would require a turbulence model with an accuracy of $\sim 99.9$\% for an 
observing time of 60\,s. This requirement would soften to an accuracy of $97-98$\% for an observing time 
of 6000\,s (provided the dynamic range increases as the square root of the observing time).

For completeness, we also computed the loss of recovered flux density of the source due to the 
turbulent atmosphere. 
In Fig. \ref{PeakRatio} we show the ratio of peak flux densities between the 
phase-referenced images and those computed without the effects from atmospheric turbulence. For 
dynamic ranges of 40-50, the loss of flux density can be as large as 25\%. The phenomenological 
model (also shown in the figure) used to fit the data is

\begin{equation}
\rho = \frac{1}{1 + D_2\,\eta^{\beta_2}}
\label{rhoNorm}
\end{equation}

\noindent where $D_2 = 0.32\pm0.02$ and $\beta_2 = 1.87\pm0.01$. 

Applicability of Eq. \ref{DmNorm} 
and \ref{rhoNorm} is not restricted to the array used in the simulations here reported. The exponents 
$\beta_1$ and $\beta_2$ only depend on the structure of the atmospheric turbulence and are thus independent 
of the interferometer used in the observations. However, the parameters $D_1$ and $D_2$ also depend on the 
stations of the interferometer. Therefore, Eq. \ref{DmNorm} and \ref{rhoNorm} can 
be adapted to any other interferometer by finding the right values of $D_1$ and $D_2$. As an example 
of this generalization of Eq. \ref{DmNorm} and \ref{rhoNorm}, Mart\'i-Vidal et al. 
(\cite{MartiVidal2010}) have studied the achievable dynamic range in phase-referenced observations with 
the Very Long Baseline Array (VLBA) at 8.4\,GHz and 15\,GHz. This study has been performed from 
quasi-simultaneous observations of 13 sources located at separations ranging from 1.5 to 20.5 degrees. 
These authors have been able to model the dynamic ranges obtained in the phase-referenced images and the 
loss of recovered flux densities (phase-referenced images compared to images obtained from self-calibrated 
visibilities) using Eq. \ref{DmNorm} and \ref{rhoNorm} with values for $D_1$ and $D_2$ different to 
those here reported, but using the same values here reported for the exponents $\beta_1$ and $\beta_2$.

\section{Conclusions}
\label{Summary}

We report on Monte Carlo estimates of the sensitivity and astrometric precision of an interferometric 
array, with a station distribution similar to that of the planned SKA, as a function of observing frequency, 
flux density, and source separation. These results can also be applied to other array distributions by 
taking into account the corresponding correction factors.
Our estimates are based on simulations of snapshot phase-referenced observations, in which we take into 
account several effects from the turbulent atmosphere and the finite temperature of the receivers. We find 
that the astrometric uncertainty strongly depends on the observing frequency and smoothly increases as the 
source separation increases. For frequencies below $\sim$1\,GHz, ionospheric effects dominate and the 
astrometry uncertainties (when the source is detectable) can be as large as $\sim$1\,as. For frequencies 
between 1 and 10\,GHz (these values slightly depend on the source flux density) atmospheric effects are 
minimum and we roughly reach the theoretical precision of the interferometer. Above these frequencies, the 
wet troposphere begins to dominate and the astrometric uncertainty increases to $\sim$10\,mas for the 
highest simulated frequency (25\,GHz). The dynamic range of the images is strongly limited by atmospheric 
turbulence at all frequencies and for all flux densities (it can decrease, in the worse cases, several 
orders of magnitude).

We propose analytical models for the loose of dynamic range and astrometric accuracy as a function of 
distance between calibrator and target source. These expressions could also be used to estimate the 
deformations and local dynamic ranges of wide-field images as a function of distance to the image phase 
center (i.e., the point in the sky where the data correlation is centered).

\begin{acknowledgements}

We thank Ed Fomalont for his very useful comments and suggestions. IMV is a fellow of the Alexander von 
Humboldt Foundation. This work has been supported by the 
European Community Framework Programme 6, Square Kilometre Array Design Studies (SKADS), contract number 
011938. This work has also been partially founded by grants Prometeo 2009/104 of the GVA and 
AYA2009-13036-CO2-2, AYA2006-14986-CO2-01, and AYA2005-08561-C03 
of the Spanish DGICYT.

\end{acknowledgements}

\appendix

\section{Complementary simulations: different number of stations and array sensitivities}
\label{OtherArrays}

Our simulations are based on a given realization of the SKA. However, the main structure of the array
distribution used in our simulations is not exclusive of the SKA. Other interferometric arrays, like
ALMA or LOFAR, are being built with similar station distributions, consisting on a compact core and
several extensions with the shape of spiral arms. Hence, our study can be extended to those arrays
by taking into account the difference between the number of stations and the station sensitivities.

The amount of noise added to the data is proportional to the number of stations, since each
station receives the signal through the turbulent atmosphere. However, for cases of clear source
detections (D $>$ 20$-$30), the dynamic range does not depend (or the dependence is weak)
on the thermal noise of the receivers (see Sect. \ref{WITHAT}). Therefore, to estimate the achievable
dynamic range for an array with a different number of stations, the
results shown in Fig. \ref{WITHATMOSPHERE} should be divided by $N/N'$, where $N$ is the number of
stations used in our simulations ($N= 200$) and $N'$ is the number of stations of the other array. This
is true for detections with a relatively large dynamic range ($D>20-30$). For weak sources, the noise
from the receivers may also contribute to the rms of the residual images, so the factor to apply in
these cases should be $(N \sigma')/(N'\sigma)$, where $\sigma$ is the thermal noise of the stations
used in our simulations and $\sigma'$ is that of the other array.

\begin{figure}[ht!]
\centering
\includegraphics[width=9cm,angle=0]{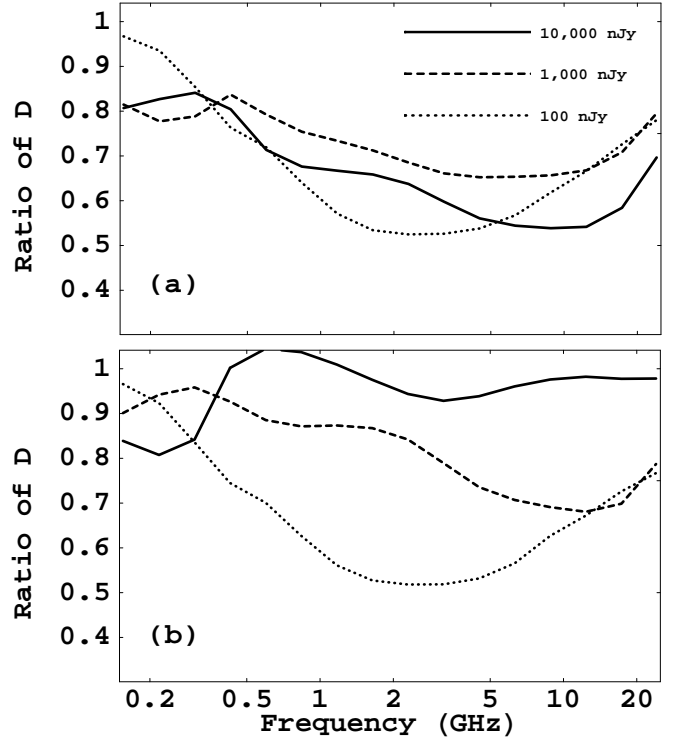}
\caption{Ratios of dynamic ranges obtained with our original array and those obtained with a subarray of
100 stations (a) and an array with half the sensitivity of the original array (b). Different lines
correspond to different target source flux densities (10\,$\mu$Jy, continuous line; 1\,$\mu$Jy dashed
line; 0.1\,$\mu$Jy, dotted line).}
\label{RATIOSARRAYS}
\end{figure}

We repeated the simulations described in Sect. \ref{WITHAT} using different arrays to compare
the results with those obtained with the original array. On the one hand, we created a smaller
array by subtracting 100 stations (those marked with empty squares in Fig. \ref{ARRAY})
from the original array. On the other hand, we created another array with all the 200 stations,
but decreasing their sensitivity by a factor 2. We show the results obtained in Fig.
\ref{RATIOSARRAYS}. Special care must be taken in the interpretation of
these figures, given that the computed ratios of dynamic ranges are only meaningful when the
detections of the sources are clear (i.e. when no spurious noise peaks appear stronger than the
source). This is true for D$>$20$-$30, which approximately corresponds to frequencies between
1 and 10\,GHz (although it slightly depends on the source flux density, see Fig.
\ref{WITHATMOSPHERE}).

With these considerations into account, we find that the ratio of dynamic ranges for the array with
100 stations falls between 0.7 and 0.5 compared to the array with 200 stations. The expected value is
0.5 (since $N' = 0.5 N$ and $\sigma' = \sigma$). Other factors, like the different coverages of Fourier
space by both arrays, could be affecting the dynamic range of the images, thus increasing the ratio
in some cases. For the array with lower station sensitivities (but the same number of stations), the
ratio of dynamic ranges falls between 0.75 and 1 for the strongest sources (as expected, since
$N' = N$ and the noise from the receivers is much smaller than the noise from the atmosphere),
while are close to 0.5 for the weakest source (also as expected, since the thermal noise from the
receivers begins to dominate in this case).

\begin{thebibliography}{.99}

\bibitem[1998]{E98} Eckers, R.D. 1999, in {\em Synthesis Imaging in Radio Astronomy II} (Taylor, Carilli 
\& Perley, eds.), ASP Conference Series Vol. 180 

\bibitem[2004]{J04} Jones, D.L. 2004, SKA Memo 45

\bibitem[2009]{MartiVidal2009} Mart\'i-Vidal, I., Guirado, J.C., Jim\'enez-Monferrer, S., \&
                               Marcaide, J.M. 2009, SKA Memo 112

\bibitem[2010]{MartiVidal2010} Mart\'i-Vidal, I., Ros, E., P\'erez-Torres, M.A., et al. 2010,
A\&A, accepted (arXiv:1003.2368)

\bibitem[2006]{Pradel} Pradel, N., Charlot, P., \& Lestrade, J.-F. 2006, A\&A, 452, 1099

\bibitem[1978]{Readhead1978} Readhead, A.C.S. \& Wilkinson, P.N. 1978, ApJ, 223, 25 

\bibitem[2007]{SAC07} Schilizzi, R.T., Alexander, P., Cordes, J.M., et al. 2007, SKA Memo 100

\bibitem[1991]{TMS91} Thomson, A.R., Moran, J.M., and Swenson, G.W., {\em Interferometry and 
Synthesis in Radio Astronomy}, 1991, Krieger Publ. Corp. (Florida)

\bibitem[2007]{vdT07} van der Tol, S., Jeffs, B.D., \& van der Veen, A.J. 2007, in 
{\em IEEE Tr. Signal Processing}

\bibitem[2009]{VLJ09} Vir Lal, D., Lobanov, A.P., \& Jim\'enez-Monferrer, S. 2009, SKA Memo 
                        (submitted)

\end{thebibliography}
\end{document}